%% file: 3c294.tex
\documentclass[usegraphicx]{mn2e}

\usepackage{amssymb}
\usepackage{mathptmx}
\usepackage{url}
\usepackage{amssymb}

\include{defn}

\begin{document}

\title{A deep \emph{Chandra} observation of the cluster environment
  of the $z = 1.786$ radio galaxy 3C294}
\author[A.C. Fabian, J.S. Sanders, C.S. Crawford and S. Ettori]
{A.C. Fabian${}^1$, J.S. Sanders${}^1$, C.S. Crawford${}^1$
  and S. Ettori${}^2$\\
${}^1$Institute of Astronomy, Madingley Road, Cambridge. CB3 0HA\\
${}^2$ ESO, Karl-Schwarzschild-Str. 2, D-85748 Garching b. Munchen, Germany}
\maketitle

\begin{abstract}
  We report the results from a 200~ks \emph{Chandra} observation of
  the $z=1.786$ radio galaxy 3C294 and its cluster environment,
  increasing by tenfold our earlier observation.  The diffuse
  emission, extending about 100 kpc around the nucleus, has a roughly
  hourglass shape in the N-S direction with surprisingly sharp edges
  to the N and S. The spectrum of the diffuse emission is well fitted
  by either a thermal model of temperature 3.5~keV and abundance $<
  0.9 \Zsun$ ($2\sigma$), or a power-law with photon index 2.3. If the
  emission is due to hot gas then the sharp edges mean that it is
  probably not in hydrostatic equilibrium.  Much of the emission is
  plausibly due to inverse Compton scattering of the Cosmic Microwave
  Background (CMB) by nonthermal electrons produced earlier by the
  radio source.  The required relativistic electrons would be of much
  lower energy and older than those responsible for the present radio
  lobes. This could account for the lack of detailed spatial
  correspondence between the X-rays and the radio emission, the axis
  of which is at a position angle of about $45^{\circ}$.  Hot gas
  would still be required to confine the relativistic plasma; the
  situation could parallel that of the radio bubbles seen as holes in
  nearby clusters, except that in 3C294 the bubbles are bright in
  X-rays owing to the extreme power in the source and the sixty fold
  increase in the energy density of the CMB.  The X-ray spectrum of
  the radio nucleus is hard, showing a reflection spectrum and iron
  line. The source is therefore an obscured radio-loud quasar.
\end{abstract}


\section{Introduction}
3C294 is a powerful Fanaroff-Riley II radio source (FR II) at a
redshift of 1.786. A high resolution image of the radio emission shows
it to have a Z-shaped morphology, which may be explained by a
precessing or torqued jet (McCarthy et al. 1990).  Extended Lyman
$\alpha$ emission was found to be associated with the source, with a
luminosity of $7.6 \times 10^{44} \ergps$ (McCarthy et al. 1990). The
source was later observed by Stockton et al. (1999) in the K$'$ band
using an adaptive optics system on the Canada-France-Hawaii Telescope,
resolving it into a number of knots arranged over an approximately
triangular region. They interpreted this pattern as a set of dusty
dwarf galaxy-like clumps illuminated in a cone by an obscured nucleus.
Quirrenbach et al. (2001) observed the system at higher resolution
with an adaptive optics system on the Keck telescope in K$'$ and H
bands, finding emission from two components with a separation of $\sim
1$~arcsec. One component was identified as the active nucleus, the
other as the core of a colliding galaxy.

X-ray emission was detected from the position of 3C294 by Crawford \&
Fabian (1996), using \emph{ROSAT} proportional counter data.  The data
quality was insufficient to conclude whether the emission was
spatially extended or thermal in origin. Their favoured model was for
X-ray emission from a rich cluster of galaxies. Hardcastle \& Worrall
(1999) later examined a long \emph{ROSAT} observation taken with its
High Resolution Imager (HRI), concluding that the source was possibly
spatially extended.

We first observed 3C294 with \emph{Chandra} in 2000, for a total of
20~ks with the ACIS-S detector (Fabian et al. 2001).  The X-ray
emission was spatially extended about the central core X-ray source in
a distinctive hourglass shape. The hourglass was aligned in the
north-south (N-S) direction and extended to radii of at least 100~kpc.
A spectrum of the diffuse emission was fitted using a thermal model,
giving a best-fitting temperature of 5~keV. The emission could have been
due to inverse Compton scattering, but since there was little excess
emission associated with the radio jet of the source, except on the
southern hotspot, this emission mechanism was considered unlikely.
The most likely explanation for the observed diffuse emission was a
thermal one, generated by the intracluster medium (ICM) of the inner
parts of a cluster.  The existence of such a hot cluster at this
redshift is consistent with a low-density universe.

Here we present the results of a ten times deeper \emph{Chandra}
exposure of 3C294, for almost 200~ks. Luminosities and distances were
calculated in this paper by assuming $H_0 = 70 \kmpspMpc$,
$\Omega_\mathrm{m} = 0.3$ and $\Omega_\Lambda = 0.7$. Relative
abundances were calculated assuming the results of Anders \& Grevesse
(1989).

\section{Data analysis}
The observation was split into two sections: one period beginning on
2002 Feb 25 with an exposure of 69.8~ks, and the second on 2002 Feb 27
with an exposure of 122.0~ks. Visual inspection of the lightcurves for
the two observations in the band 0.3 to 5~keV showed no evidence for
contamination by flares, therefore the total effective length of
observations was 191.8~ks.

Both of the data sets were taken using the ACIS-S3 detector in VFAINT
mode. This mode of observation grades events using a $5\times 5$
matrix of detector pixels around the centroid of each event, rather
than the standard $3 \times 3$ matrix for the FAINT observation mode,
and yields a dataset with a substantially lower background.

In order to take advantage of the VFAINT observation mode, the data
were reprocessed using the CAIO \textsc{acis\_process\_events} tool
with the check\_vf\_pha option switched on.  The gain file applied to
the reprocessed dataset was acisD2000-08-12gainN0003.fits from CALDB
version 2.12.

Since the two data sets were taken over only a few days, the roll
angles of the spacecraft were within a couple of degrees of each
other.  Therefore we merged the two events files together into a
single file, reprojecting the position of the events (in sky aligned
detector coordinates) of the first data set to match that of the
second. To test whether this step was valid we generated
spatially-weighted response and ancillary matrices for the diffuse
emission for the two data sets individually and compared the
differences in fitting the spectra using the two responses on both
data sets. There was no significant difference in either the quality
of the fits or their best-fitting parameters.

The weighted responses for each region were created from the
appropriate FEF calibration file (acisD2000-01-29fef\_piN0002) and the
CIAO \textsc{mkwarf} and \textsc{mkrmf} tools, weighting the response
using the number of counts in the 0.3 to 7.5~keV band.

Creating a suitable background spectrum is an important part of the
analysis of low surface brightness objects such as the diffuse
emission here. Given that the emission is small in extent and the
observation is long, it is preferable to use a background spectrum
generated from the observation itself rather than different
observations.  A blank-sky background dataset would have a different
instrument response and galactic absorption to our observation.  We
therefore used blank regions of the S3 detector in the merged events
file to provide the background. Too small a region gives uncertain
background subtraction, and too large risks introducing systematics
from the variation of the background over the chip. We used a
background field $4.97 \times 1.67 \aminsq$ along the CCD node which
harbours the diffuse emission, subtracting a circle of radius
$0.41\amin$ centred on the central source.

\subsection{The diffuse emission}
\label{sec:analysis_diffuse}
\begin{figure}
  \centering
  \includegraphics[width=0.8\columnwidth]{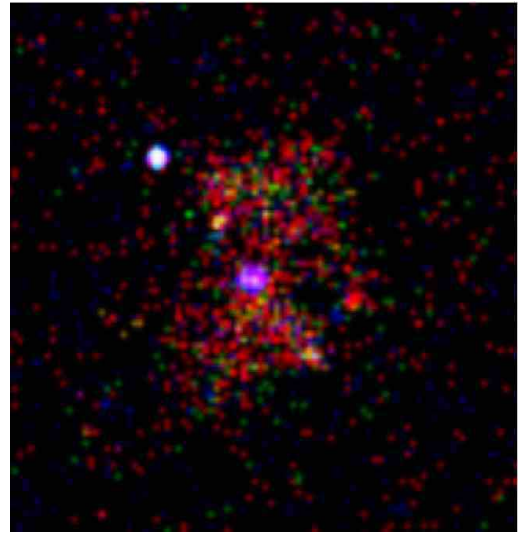}
  \caption{Energy-mapped image of the emission. Events between 0.3 to
    1~keV are coloured red, 1 and 1.5 keV green, and 1.5 to 5~keV
    blue. The data are plotted with 0.49~arcsec pixels, smoothed with
    a Gaussian of width 0.25~arcsec. The size of the image is $47
    \times 48 \asecsq$, corresponding to an area of roughly $400
    \times 400 \kpcsq$ at the redshift of the radio galaxy. In this
    and all our images, north is to the top and east is to the left.}
  \label{fig:smooth_rgb}
\end{figure}

\begin{figure}
  \centering
  \includegraphics[width=0.8\columnwidth]{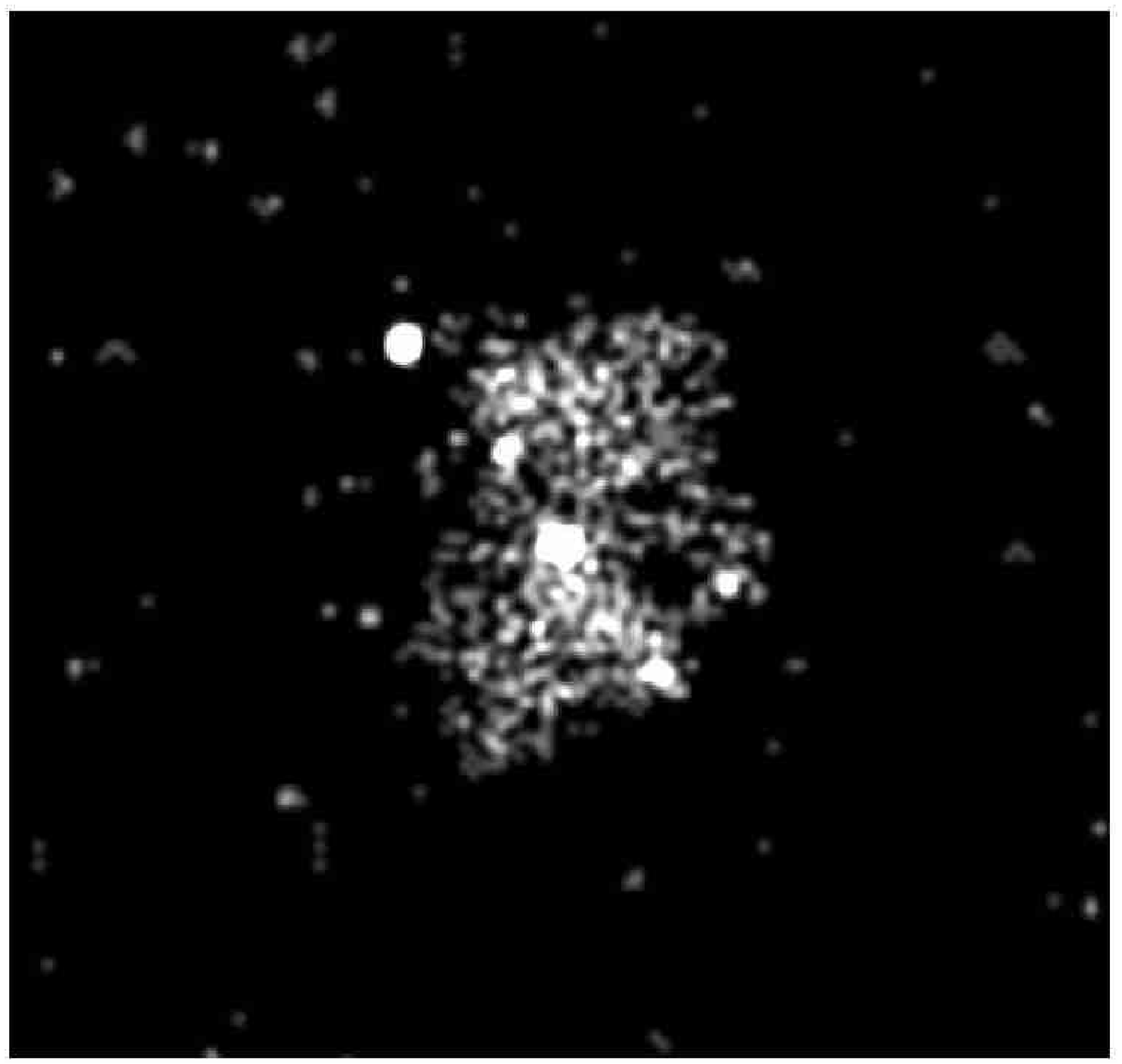} \\
  \includegraphics[width=0.8\columnwidth]{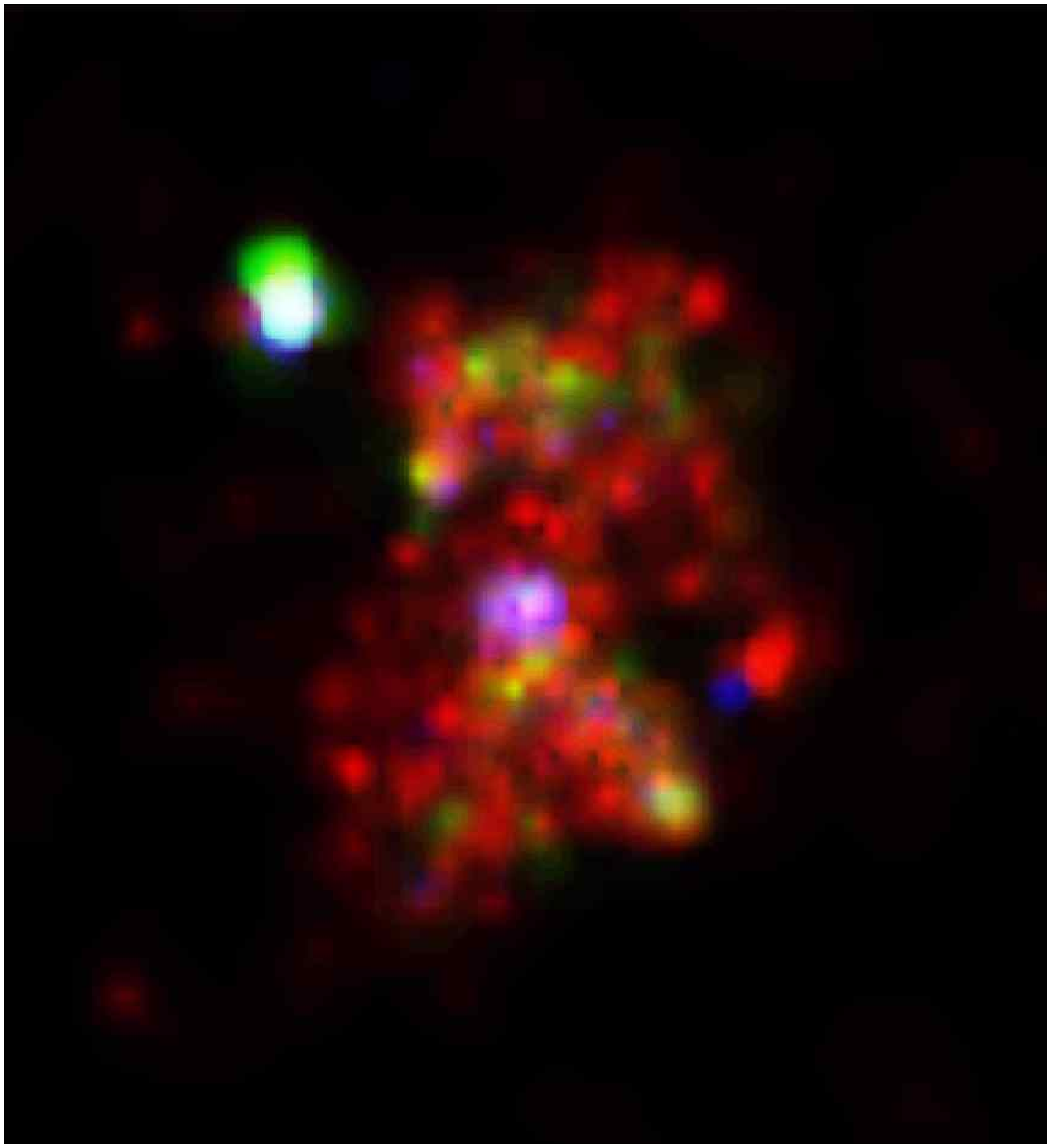}
  \caption{Greyscale image of the diffuse emission between 0.3 and
    5~keV using a background removal technique and smoothed with a
    Gaussian of width 0.5~arcsec (top). Energy mapped image of the
    diffuse emission (as in Fig. \ref{fig:smooth_rgb}) smoothed using
    a tessellation technique (bottom).}
  \label{fig:diff_best}
\end{figure}

\begin{figure*}
  \centering
  \includegraphics[width=0.99\textwidth]{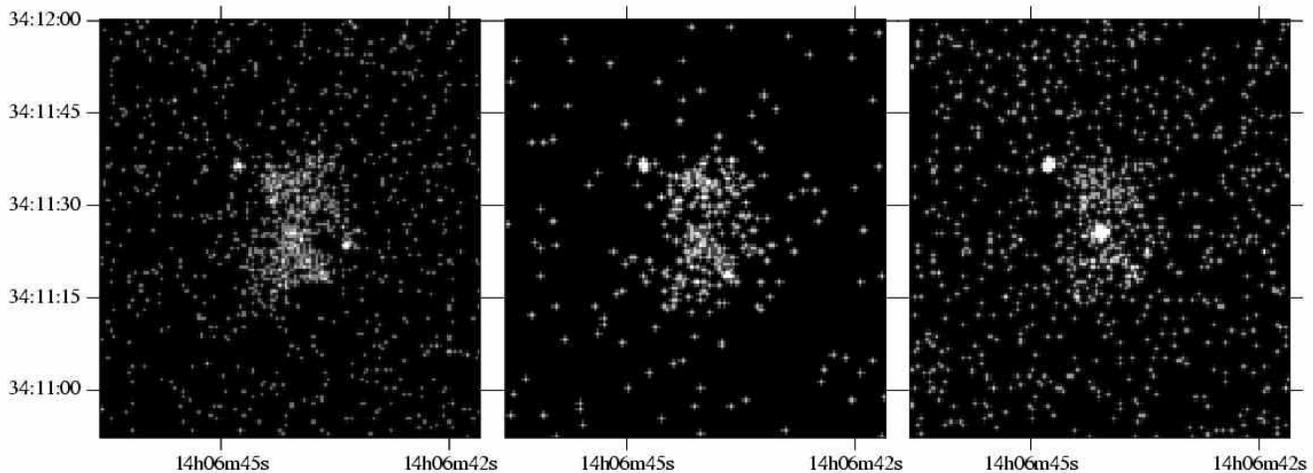}
  \caption{Images of the emission in the 0.3-1.0 (left), 1.0-1.5
    (middle) and 1.5-5.0~keV (right) bands. Data are shown with
    0.49~arcsec pixels, smoothed with a Gaussian of width
    0.25~arcsec.}
  \label{fig:threebands}
\end{figure*}

\begin{table}
  \centering
  \begin{tabular}{lll}
    Region & Fit-type & Best-fitting parameters and reduced $\chi^2$\\ \hline

    Core & Power-law & $\Gamma =-0.65$, $\chi^2_{\nu} = 2.7$ \vspace{1mm}\\

    & \textsc{pexrav}${}^{*}$ & rel. refl$=5.1^{+2.0}_{-1.5}$,\\
    &&cover. $N_H=8.4^{+1.1}_{-0.9} \times 10^{23} \psqcm$, \\
    && cover. fract.$=97.6^{+1.0}_{-3.0}$\%, $\chi^2_{\nu} = 0.55$
    \vspace{2mm} \\

    NE & Power-law & $\Gamma=0.7$, $\chi^2_{\nu} = 0.77$  \vspace{1mm}\\

    & Par. Cover.${}^{\dagger}$ & cover. $N_H=2.5_{-1.7}^{+1.4} \times 10^{22} \psqcm$,\\
    &&cover. fract.$=75_{-56}^{+14}$\%, $\Gamma=1.6^{+1.1}_{-1.0}$,\\
    &&$\chi^2_{\nu} =0.82$ \vspace{2mm}\\

    Diffuse & \textsc{mekal} &
    $\mathrm{k}T = 5.9^{+1.8}_{-1.3}$~keV, $Z = 0.07^{+0.26}_{-0.07}$, \\
    &&$N_H = 3.8^{+2.1}_{-1.8} \times 10^{20}\psqcm$,
    $\chi^2_{\nu} = 1.0$ \vspace{1mm} \\

    & Power-law & $\Gamma =2.3^{+0.3}_{-0.1}$,\\
    &&$N_H=1.1\pm0.3 \times 10^{21} \psqcm$,
    $\chi^2_\nu = 0.91$ \vspace{1mm} \\

    & \textsc{mekal}${}^\ddagger$ & $\mathrm{k}T =
    3.5^{+0.6}_{-0.5}$~keV, $Z = 0.30^{+0.30}_{-0.25} \Zsun$,\\
    &&$\chi^2_\nu = 0.98$ \vspace{1mm} \\

    & Power-law${}^\ddagger$ & $\Gamma=2.29 \pm 0.10$,
    $\chi^2_\nu = 0.64$ \\

    \hline
  \end{tabular}
  \caption{Summary of results of spectral fits. Uncertainties shown
    are $1\sigma$. ${}^*$\textsc{pexrav} reflection
    disc model plus emission from neutral iron line, absorbed with
    partial covering model. ${}^\dagger$Power-law model absorbed with
    a partial coverer. ${}^\ddagger$\textsc{mekal} or power-law model
    absorbed with \textsc{acisabs} model plus \textsc{phabs} model set
    to Galactic absorption.}
  \label{tab:fits}
\end{table}

The diffuse X-ray emission has the distinctive hourglass shape seen in
our earlier observation of this object (Fabian et al. 2001). In
Fig.~\ref{fig:smooth_rgb} we show images of the emission, smoothed by
a Gaussian, in three observed energy bands (0.3-1.0, 1.0-1.5 and
1.5-5.0 keV), superimposed as red, green and blue.

In order to see the diffuse emission above the background easily we
have applied a background-reduction technique (analogous to applying a
cut in brightness to an image with many photons), the result of which
shown in Fig.  \ref{fig:diff_best} (top). We removed those photons
which did not have 3 or more neighbouring photons within a radius of
1~arcsec from them.  The removed photons were in a flat distribution
around the diffuse emission.

To highlight the main features of the three-colour image (Fig.
\ref{fig:smooth_rgb}) we show a smoothed version of it in Fig.
\ref{fig:diff_best} (bottom). We used the bin accretion algorithm of
Cappellari \& Copin (2002) to create a tessellated image in each of
the three bands with a ratio of signal to Poisson noise of 0.8. The
centroids of the cells formed by tessellation were then interpolated
with the \textsc{natgrid} natural neighbour interpolation
library\footnote{http://ngwww.ucar.edu/ngdoc/ng/ngmath/natgrid/nnhome.html}
to form images, which were superimposed as red, green and blue layers.

The raw images are shown individually in Fig.~\ref{fig:threebands},
emphasizing that the emission is extended in each of the three bands.
The diffuse emission is still visible in a 3-5~keV image,
corresponding to 8.4-13.9~keV in the rest frame. Binning the data into
16~arcsec bins shows there is no evidence for any low intensity
emission surrounding the central diffuse emission (we calculate limits
for the outlying gas in Section~\ref{sect:hydrostatic}).

The images also show several point sources: a central hard source
corresponding to the position of the radio core, another hard source
14~arcsec to the north-east (NE), a soft source 9~arcsec to the
west-south-west (WSW), and two fairly soft sources along the direction
of the radio axis (6~arcsec to the NE and 8~arcsec to the SW)
corresponding to the position of the radio hotspots. We will discuss
these point sources in later sections.

\begin{figure}
  \centering
  \includegraphics[width=0.99\columnwidth]{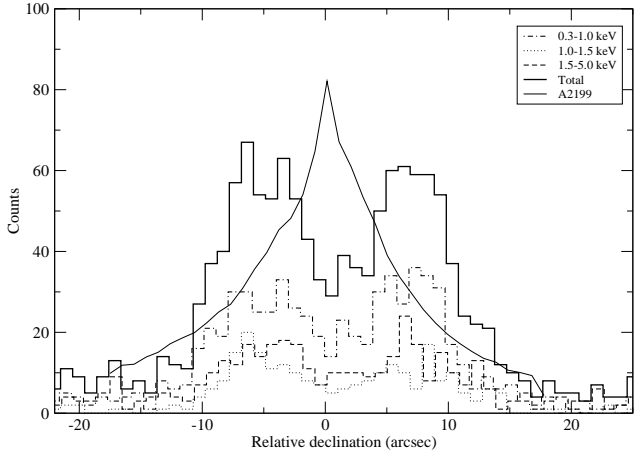}
  \caption{Surface brightness profile along the north-south direction
    (negative values are south), excluding the core, NE source and the
    WSW source. Values shown are the number of counts in a $0.98\times
    20.6$~arcsec${}^2$ box, in the bands 0.3-1, 1-1.5, 1.5-5.0 and
    0.3-5~keV. The background level is around 5 counts in the 0.3-5
    keV band (i.e. beyond 15-20 arcsec is \emph{just} background). The
    continuous peaked line shows a similar north-south profile of
    A2199, on the same spatial scale as 3C294, in a band equivalent to
    0.3-1.0~keV.  The counts in the central 3 bins are underestimated
    since the core source was removed by excluding a 1.6~arcsec radius
    circle.}
  \label{fig:surbrightnorth}
\end{figure}

The surface brightness declines sharply at the edge of the hourglass
shape.  Fig.~\ref{fig:surbrightnorth} shows a surface brightness cut
in a north-south direction, measured by moving a $\sim 21 \times 1
\aminsq$ box along that axis, excluding the point source in the
centre, the NE and the WSW. We show the number of counts in the three
observed bands: 0.3-1.0, 1.0-1.5 and 1.5-5.0~keV. The profile appears
box-like with the number of counts dropping from about 60 to 10 in
around 5~arcsec at the N and S edges.

\begin{figure}
  \centering
  \includegraphics[width=0.8\columnwidth]{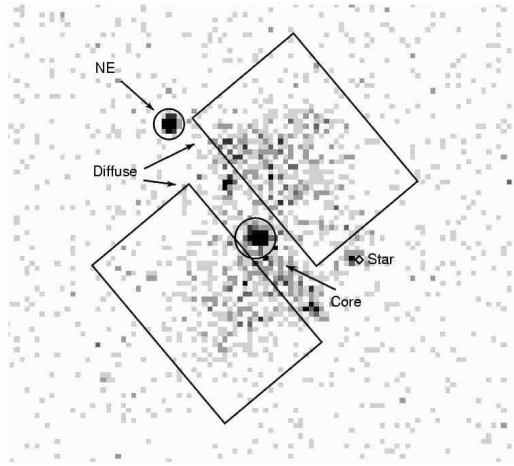}
  \caption{Regions used for spectral extraction, overlayed on the raw
    X-ray image of the diffuse emission in the 0.3-5.0~keV band. The
    two boxes were the regions used for the analysis of the diffuse
    emission.  The circle to the NE was the NE source region, and the
    circle in the centre was the region used for the core. The diamond
    on the right marks the USNO-A2.0 position of the star.}
  \label{fig:regions}
\end{figure}

\begin{figure}
  \centering
  \includegraphics[angle=-90,width=0.99\columnwidth]{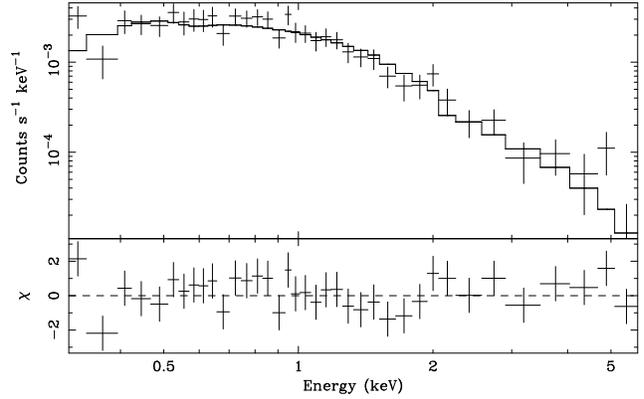}
  \caption{Spectrum of the diffuse emission between 0.3 and 6~keV
    (from the boxed region in Fig.\ref{fig:regions}, also showing
    $\chi$ contribution from each spectral bin when fitted with a
    single-component \textsc{mekal} model.}
  \label{fig:clustspec}
\end{figure}

The profile in the east-west direction declines more smoothly than
that in the north-south direction. In the southernmost lobe the
diffuse emission declines slowly to the east. In the northernmost lobe
it declines slowly to the west.

We fitted the 0.3-6 keV spectrum of the diffuse emission from the
regions shown in Fig.~\ref{fig:regions} by a \textsc{mekal} model
(Mewe, Gronenschild \& van den Oord 1985; Liedahl, Osterheld \&
Goldstein 1995) with a \textsc{phabs} absorbing screen
(Balucinska-Church \& McCammon 1992), giving best-fitting parameters
of $\mathrm{k}T = 5.9^{+1.8}_{-1.3}$~keV, $Z = 0.07^{+0.26}_{-0.07}
\Zsun$, $N_H = 3.8^{+2.1}_{-1.8} \times 10^{20}\psqcm$ (with $1\sigma$
uncertainties).  At 2$\sigma$, the uncertainty on the temperature was
$\mathrm{k}T = 5.9^{+3.4}_{-2.0}$~keV. The data were binned into
spectral bins containing at least 20 counts. The spectrum is shown in
Fig.~\ref{fig:clustspec}, with the contributions to $\chi^2$ from each
spectral bin. The reduced $\chi^2$ of the fit was 1.00 (33.0/33). The
regions were chosen to exclude the emission from along the radio axis
of the source, an area which could be contaminated by inverse Compton
emission. We summarise the results of spectral fits in
Table~\ref{tab:fits}.

We also fitted the spectrum of the diffuse emission using a power-law
absorbed by a \textsc{phabs} model. This model fitted the spectrum
well (reduced $\chi^2 = 0.91 = 31.0/34$) with a photon index of
$2.3_{-0.1}^{+0.3}$, however it required strong absorption of $(1.1
\pm 0.3) \times 10^{21} \pcmsq$ which would be unreasonable if it were
distributed over the whole volume of the diffuse emission.

Although both models fitted so far require excess absorption, there is
a known degradation of the ACIS soft X-ray
response\footnote{http://cxc.harvard.edu/cal/Links/Acis/acis/Cal\_prods/qeDeg/}
which may account for this. We have refitted the spectra making use of
the preliminary response model \textsc{acisabs} of Chartas \& Getman
(2002). Fitting the spectrum from 0.4-6~keV by a power-law model
(increasing the lower energy bound of the spectral fit as suggested by
the documentation of this model) and absorbing it with the
\textsc{acisabs} model plus a \textsc{phabs} model (fixed to the
Galactic value of $1.2 \times 10^{20} \pcmsq$), the reduced $\chi^2$
of the best-fit was $0.64=20.6/32$.  The best-fitting photon index was
$2.29 \pm 0.10$ ($1\sigma$), and the intrinsic luminosity of the
power-law was $2.5 \times 10^{44} \ergps$ between 1 and 10~keV in the
restframe.  There was no indication of an iron line in the residuals
of the power-law fit. We note, however, that the energy of an iron
line would lie close to the iridium M-edge of the mirror, which could
cause a problem if the spectral calibration is not perfect.

We also fitted a \textsc{mekal} model absorbed by the \textsc{acisabs}
and \textsc{phabs} models as above, allowing the abundance and
temperature to be free. In that case the best-fitting temperature of
the gas was $3.5_{-0.5}^{+0.6}$~keV ($1\sigma$) or
$3.5_{-0.7}^{+1.1}$~keV ($2\sigma$). The best-fitting abundance was
$0.30_{-0.25}^{+0.30} \Zsun$ ($1\sigma$) or $0.30_{-0.30}^{+0.57}
\Zsun$ ($2\sigma$). A plot of $\chi^2$ space shows the errors in
temperature and abundance to be largely orthogonal. The
reduced $\chi^2$ of the best-fitting parameters was $0.98 = 30.3/31$.
Separate spectra of the northern and southern lobe were made, and
fitted simultaneously with the absorbed \textsc{mekal} model above.
There was no evidence in the residuals of the fit indicating that the
spectra of the two lobes were different.

As a final check, we refitted the spectrum of the diffuse emission by
minimizing a modified version of the C-statistic (allowing background
subtraction; Arnaud 2002) rather than the $\chi^2$-statistic. With the
\textsc{mekal} model, we found temperature and abundance of
$kT=3.9^{-0.5}_{+0.6}$ and $Z=0.2_{-0.2}^{+0.2} \Zsun$ ($1\sigma$),
fitting the spectrum from 0.4-6~keV.  Fitting from 1-6~keV, we
obtained a temperature of $7.4_{-1.6}^{+2.7}$~keV, confirming the
result above using the $\chi^2$-statistic.  A power-law model gave the
best-fitting spectral index $2.26^{+0.09}_{-0.10}$. The C-statistic
does not provide a goodness-of-fit.

In summary, providing that we assume that the \textsc{acisabs} model
is appropriate for correcting the ACIS-S low energy response, both
\textsc{mekal} and power-law models give acceptable fits to the
spectrum of the diffuse emission. The best-fitting \textsc{mekal}
temperature is $\sim 3.5$~keV and the abundance is $<0.9\Zsun$
($2\sigma$), with a reduced $\chi^2$ of 0.98. The best-fitting
power-law spectral index is $2.3$, with a reduced $\chi^2$ of 0.64.

\subsection{The source to the West-South-West (WSW)}
\begin{figure}
  \centering
  \includegraphics[width=0.99\columnwidth]{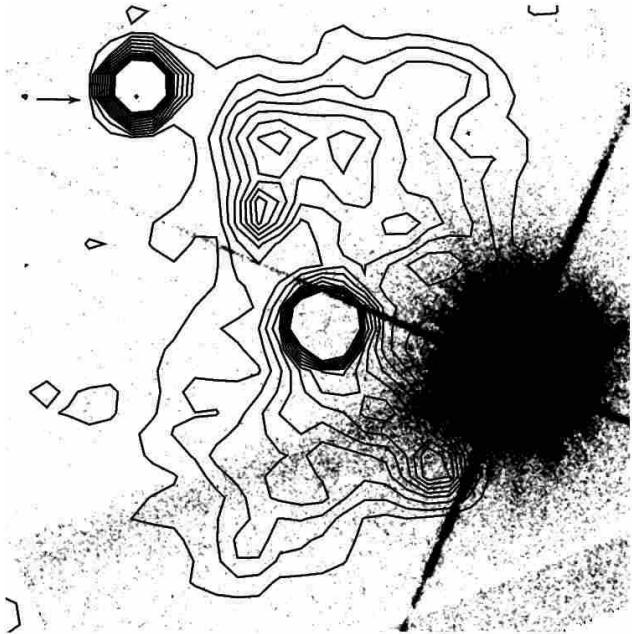}
  \caption{Smoothed \emph{HST} image of the area of the diffuse X-ray emission,
    overlayed by contours from a smoothed X-ray image, adjusted so
    that the optical position of the star is coincident with the
    position of its X-ray source.  North is to the top in this image.
    The arrow shows the position of a very faint optical point source
    coincident with the NE X-ray source.  There is an optical
    elongated structure at the position of the X-ray nucleus of
    3C294.}
  \label{fig:hst}
\end{figure}

We extracted archival \emph{Hubble Space Telescope} (\emph{HST}) Wide
Field Planetary Camera 2 (WFPC2) data of four observations of 3C294
(datasets U27LFC0\{1,2,3,4\}T). The observations were taken using a
filter with a central wavelength of 6895\AA, and each had an exposure
time of 140~s. The data were combined using the \textsc{iraf} task
\textsc{crrej}.

The \emph{HST} field in Fig. \ref{fig:hst} is dominated by a star
known as U1200-07227692 in the USNO-A2.0 catalogue (Monet et al.
1998). The star has B and R-band magnitudes of 13.0 and 11.5 in that
catalogue, and a V band magnitude of 12.0 (Hubble Guide Star Catalogue
1.2, Lasker et al. 1990). Quirrenbach et al. (2001) discuss the
various disagreeing measurements of the position of this star, but
state that the USNO-A2.0 position is the most accurate. The position
in this catalogue (14:06:43.3, +34:11:23.5) matches within an arcsec
the position of the soft X-ray source to the south-east-east of the
central nucleus (Figs. \ref{fig:threebands} \& \ref{fig:regions}).

This star was reported to be an F-type star by Wyndham (1966; spectrum
taken by F.~Zwicky 1965). Stockton, Canalizo \& Ridgway (1999) found
the star to be a double with a separation of 0.13~arcsec, and an
intensity ratio of 1.5:1 in $K'$. McCarthy et al. (1990) from its
spectrum find it to be a subgiant K star, with a spectral break at
4000\AA.

The flux of this source is $\sim 6 \times 10^{-16}\ergpcmsqps$ between
0.3 and 5~keV. This flux is low (relative to its optical luminosity)
when compared with the fluxes of optically bright main-sequence stars
(H\"unsch, Schmitt \& Voges 1998). The low flux is more typical of
subgiant K stars than F-type stars.

\subsection{The North-East (NE) source}
\label{sect:ne}
\begin{figure}
  \centering
  \includegraphics[angle=-90,width=0.99\columnwidth]{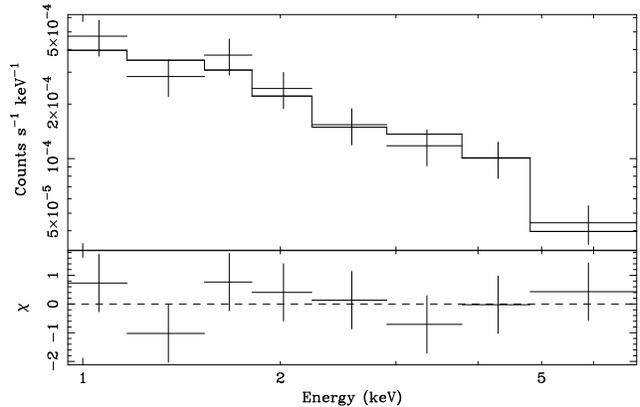}
  \caption{Spectrum of the source to the NE of the radio
    source. The spectral fit shown is a power law with a partial
    covering model. Data are binned into spectral bins containing at
    least 20 counts.}
  \label{fig:nespec}
\end{figure}

The X-ray spectrum of the source to the NE of the core is shown in
Fig.  \ref{fig:nespec}. It is well fitted by an absorbed
($N_\mathrm{H} \sim 10^{20}$\psqcm) power-law model, but the
best-fitting photon index is unphysical at 0.7. A partial covering
model fits better, with an absorption of $2.5_{-1.7}^{+1.4} \times
10^{22} \psqcm$, a partial covering fraction of $75_{-56}^{+14}$~per
cent, a power law index of $1.6_{-1.0}^{+1.1}$ ($1\sigma$ errors). The
reduced $\chi^2$ for the fit was $0.82 = 3.3/4$. The model was fitted
to the data in spectral bins of at least 20 counts, and between the
energies of 0.5 and 7~keV. We added Galactic absorption and corrected
the data with the \textsc{acisabs} model, although the correction
makes no difference to the best-fitting parameters.  The source
contains very few counts below 0.8~keV, suggesting it is heavily
absorbed. If this source lies at the same redshift as 3C294 its
intrinsic luminosity is $2.8 \times 10^{44} \ergps$ between 1 and
10~keV.

The \emph{HST} image of 3C294 in Fig.~\ref{fig:hst} contains a
very-faint point source at the location of the NE source. We estimated
a photometric redshift for this source using the publicly available
code \textsc{hyperz} (Bolzonella, Miralles \& Pell\'{o} 2000), and
calibrated, archival imaging data from the HST (F702 filter), INT
Wide-Field Camera (i\'\ Sloan filter) and UKIRT (K filter). The best
solution is consistent with $z=z_\mathrm{3C294}$, though we emphasize
that this is based on three filters only. The best-fitting solution is
$z=1.93_{-1.23}^{+0.14}$, and there is a secondary solution
$z=2.97_{-0.97}^{+0.41}$ (90 per cent confidence intervals).

The flux from this source in X-rays varied between the original
\emph{Chandra} observation and the one we present here. In the
original observation on 2000 October 29 the count rate between
0.5-7~keV was $(2.2 \pm 0.3) \times 10^{-3} \ps$.  The combined count
rate from the merged datasets here is $(9.2 \pm 0.7) \times 10^{-4}
\ps$. Therefore the flux of the source declined by a factor of $\sim
2.4$ in $4.2 \times 10^7$~s.

\subsection{The core spectrum}
\label{sect:core}
\begin{figure}
  \centering
  \includegraphics[angle=-90,width=0.99\columnwidth]{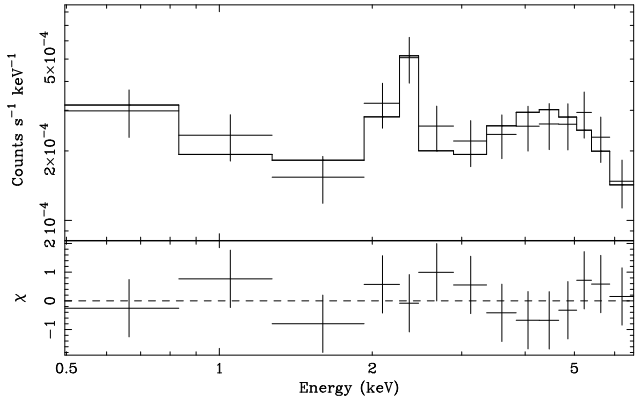}
  \caption{Spectrum and \textsc{pexrav} plus iron line fit of the central X-ray
    source. Data are binned into spectral bins containing at least 20
    counts and fitted between 0.5 and 7~keV.}
  \label{fig:corespec}
\end{figure}

\begin{figure}
  \centering
  \includegraphics[angle=-90,width=0.99\columnwidth]{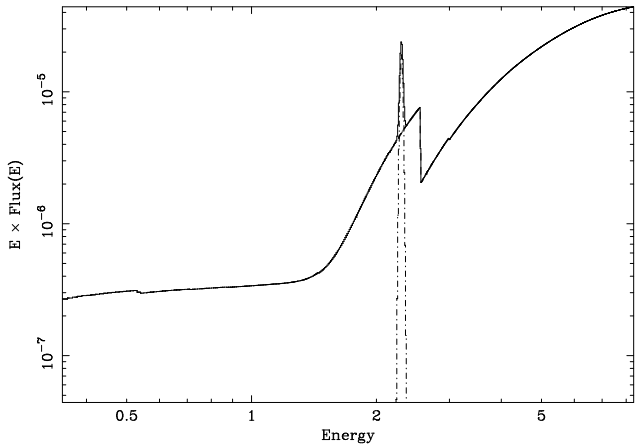}
  \caption{Best-fitting model for the core spectrum, plotted as $E \:
    F_E$. The width of the iron line is set here to 0.05~keV. Model
    includes partial coverer, \textsc{acisabs} and Galactic
    absorptions.}
  \label{fig:coremodel}
\end{figure}

A power law model is not a good fit to the spectrum of the core
source, associated with the central active nucleus. The best-fitting
photon index was not physical at $-0.65$, and the reduced $\chi^2$ was
2.7. The data were fit between 0.5 and 7 keV.

A much better fit is found using a partially-obscured \textsc{pexrav}
(Magdziarz \& Zdziarski 1995) reflection model plus emission from a
neutral iron line at 6.4~keV (Fig.  \ref{fig:corespec}), with Galactic
absorption.  \textsc{acisabs} correction had little effect on the best
fitting values, although the uncertainties on the best-fitting
parameters were lower since we did not fit for Galactic absorption,
instead fixing it. We quote the values corrected by \textsc{acisabs}
absorption here. The partial coverer was placed at the redshift of the
source.

The relative ratio of the reflected to direct component was found to
be $5.1_{-1.5}^{+2.0}$ ($1\sigma$). The covering model had a column
density of $8.4^{+1.1}_{-0.9} \times 10^{23} \psqcm$ and a covering
fraction of $97.6^{+1.0}_{-3.0}$~per~cent ($1\sigma$). Solar abundance
was assumed, as was an inclination angle of $60^{\circ}$, and the
width of the iron line was set to zero. The reduced $\chi^2$ of the
fit was $0.55 = 5.0/9$. The flux in the iron line was
$2.0_{-1.1}^{+1.3} \times 10^{-6}$ photon\ps\psqcm. Using an F-test,
we estimate there is only a 3.5 per~cent probability of an intrinsic
spectrum without the added neutral iron line giving as a good a fit as
a model with the line.

Fig.  \ref{fig:coremodel} shows the spectrum of the best-fitting
model, with the width of the iron line set to 0.05~keV.  The intrinsic
luminosity of the underlying power law is $1.1 \times 10^{45} \ergps$
between 1 and 10~keV in the rest frame, after correction for
absorption. We note that the direct emission leaking to our line of
sight below 1.5~keV (Fig. \ref{fig:coremodel}) need not be due to
partial covering, but just to some of the radiation being scattered by
ionized gas. In this case the above nucleus luminosity is a lower
limit.

The \emph{HST} image in Fig.~\ref{fig:hst} shows a faint elongated
structure at the position of the X-ray nucleus, aligned roughly in the
north-south direction.

\subsection{Radio hotspots}
\label{sect:radio_hotspots}
\begin{figure}
  \centering
  \includegraphics[width=0.99\columnwidth]{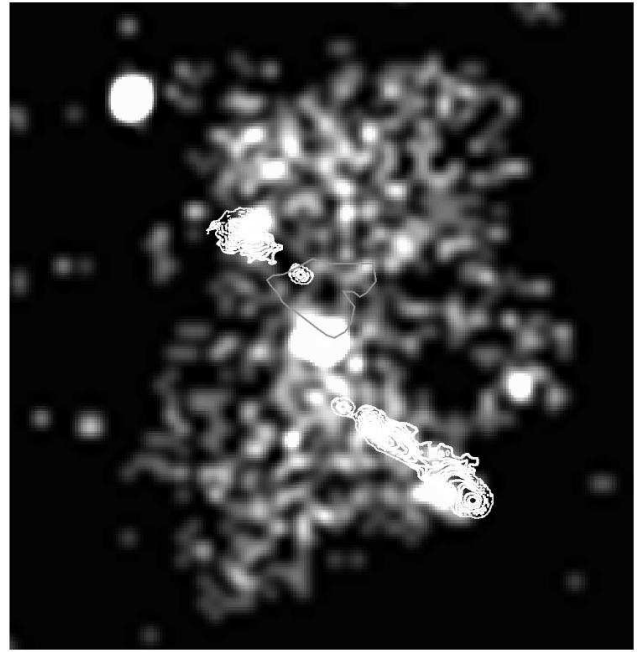}
  \caption{X-ray emission between 0.3 and 5 keV shown in greyscale
    (as in Fig. \ref{fig:diff_best} [top]).  Overlayed are the 6~cm
    radio emission contours, and a Lyman-$\alpha$ contour (roughly
    triangular) taken from Fig.~2, McCarthy et al.  (1990).}
  \label{fig:radio}
\end{figure}

Fig. \ref{fig:radio} shows an overlay of the X-ray emission between
0.3 and 5~keV, overlayed with 6~cm radio and a Lyman-$\alpha$ contour
taken from Fig.~2 of McCarthy et al. (1990). To make this image, the
McCarthy et al. data were converted to J2000 coordinates, and the
central radio source was aligned with the \emph{Chandra} X-ray source.
The radio hotspots are in a similar position to the two X-ray point
sources along the radio axis, but the X-ray sources are rotated by
about 6 degrees clockwise about the core from the radio hotspots, and
appear to be closer in.  Each spot is displaced by about 1.8~arcsec.
The X-ray sources contain too few counts to generate a spectrum, but
can easily be seen in the 0.3-1 and 1-1.5~keV images, but not from
1.5-5~keV, indicating that they have soft spectra ($\Gamma > 2.3$, the
value of the diffuse flux, Section \ref{sec:analysis_diffuse}). If we
assume a photon index for the spectrum of the hotspots of 2.3, the
unabsorbed flux emitted in a 1.2~arcsec radius circle about the NE
hotspot from 1-10~keV is $5.8 \times 10^{-16} \ergpcmsqps$ (rest frame
1-10~keV luminosity $L_{1-10} \sim 1.7 \times 10^{43} \ergps$). The
intrinsic flux for the SW hotspot is $5.5 \times 10^{-16} \ergpcmsqps$
($L_{1-10} \sim 1.6 \times 10^{43} \ergps$).

\section{Interpretation of the point sources}
\subsection{The NE source}
This source is bright in the X-ray band and has a low optical flux.
The X-ray spectrum indicates that it is highly absorbed. The source
has also varied in X-ray brightness strongly over 18 months. It is
therefore likely to be a Seyfert II galaxy (of an intrinsic luminosity
of $2.5\times 10^{44}\ergps$ between 1 and 10~keV), harbouring a
highly obscured AGN. The galaxy may be a member of the cluster, a
possibility supported by its photometric redshift.

\subsection{Radio axis features}
Four sources appear to lie along the direction of the radio axis: the
core (\S\ref{sect:core}), two sources coincident with the radio hot
spots, and the NE source (\S\ref{sect:ne}). Indeed, the NE source is
on a direct line joining the lower hotspot and the core. It is
tempting to directly connect the NE source with the radio axis, but it
is difficult to suggest a mechanism describing how the radio axis
could trigger activity in the NE source, which appears to be a Seyfert
II.  The hotspots, the core, and the NE source are probably aligned by
chance.

There is an enhancement of the X-ray emission along the SW side of the
jet, and there is also a hole in emission along the NE side (Figs.
\ref{fig:radio} and \ref{fig:diff_best}). This may indicate that the
jet is relativistic and beamed, with the SW side being beamed towards
us.

\subsection{The central source}
The spectrum of the central source is fit well with a model of a
highly obscured, reflection-dominated, AGN.  The intrinsic luminosity
of the nucleus is of the order of $10^{45} \ergps$.

\section{Interpretation of the diffuse emission}

\subsection{Thermal}

\subsubsection{Hydrostatic models}
\label{sect:hydrostatic}
If we approximate the emitting volumes as two spheres of radius
6.4~arcsec, then the total emitting volume is $3.9\times
10^{70}\cmcu$. Using the normalisation of the \textsc{mekal} model
(with the \textsc{acisabs} correction), we find the electron density
to be approximately $2.6\times 10^{-2} \pcmcu$. The energy content of
the hot gas is therefore of the order $(3/2) k_B T n V \sim 1.5 \times
10^{61} \erg$ (assuming $kT\sim 3.5$~keV), where $n$ is the total
particle density. If this were delivered over $10^8 \yr$, this would
require a power of $5 \times 10^{45} \ergps$, not including
$P\mathrm{d}V$ work.

If we are observing gas in hydrostatic equilibrium, the sharp edges at
100~kpc mean that the scale height there is $\sim 20$~kpc. The virial
temperature of the cluster then has to be $\sim 100/20 = 5$ times the
temperature of the gas. Therefore if this is a cluster in hydrostatic
equilibrium it must be very hot, extremely massive, and more massive
than RX~J1347-1145 (Allen, Schmidt \& Fabian 2002).

Sharp X-ray edges are seen in many nearby clusters in a phenomenon
known as a cold front (Markevitch et al. 2000). The gas temperature
decreases sharply across these features, with the cooler gas being
closer to the centre; pressure is continuous across the front. If this
is the explanation for the sharp edges to the diffuse X-ray emission
of 3C294, there must be hotter surrounding gas.  Assuming pressure
equilibrium across the front, then the ratio of the temperatures of
the gas either side of the front is approximately equal to the inverse
of the ratio of their densities. If we estimate the density drop as a
factor of 2.5 (estimated from Fig. \ref{fig:surbrightnorth}), then the
outer temperature should be roughly 8~keV. This again makes the
cluster very hot and rare at its redshift (Fabian et al. 2001).
Moreover, the interiors of nearby clusters with cold fronts do not
show brightness profiles as flat as we see here.

We can place a limit on the luminosity of any gas that lies outside of
the observed diffuse emission. We calculated the flux limit in a shell
between radii of 16 and 55~arcsec from the core, assuming a temperature
for the gas of 3.5~keV, an abundance of $0.3\Zsun$, and Galactic
absorption, and find the $3\sigma$ upper limit on the intrinsic
luminosity of that gas between 1 and 10~keV to be $\sim 10^{44}
\ergps$. Using this limit we can say that the mean density of gas in
the shell must be less than $1/5$ of the average density of gas making
the observed extended X-ray source. This is inconsistent with the cold
front hypothesis.

Fabian et al. (2001) concluded that the diffuse emission found around
the point source in 3C294 was most likely direct X-ray emission from
the core of a cluster.  Unfortunately our analysis of this longer
observation presents some difficulty for this interpretation. In
particular, the sharp drops observed at the edges of the diffuse
emission are in contrast to the expected profile from intracluster
gas. We plot a surface brightness cut of Abell~2199 (taken from data
in Johnstone et al.  2002), a cluster at a redshift of 0.0309, on
Fig.~\ref{fig:surbrightnorth}.  This cut was produced by enlarging the
size of the moving box by a factor of 13.7, corresponding to the
relative angular diameter distance of 3C294 compared to A2199, and
examining an energy range of 0.82 to 2.71~keV, corresponding to an
energy range of 0.3 to 1~keV at the redshift of 3C294.  The cut
through A2199 has a large central peak, indicating gas with a short
cooling time in the centre, and no sharp drops in brightness. The cut
through 3C294 is quite unlike A2199, with a largely flat profile and
sharp drops at the edges. Note, however, that the diffuse emission has
a less abrupt edge to the east and west. The sharpness of the N and S
edges to the X-ray emission mean that the gas is disturbed and not in
complete hydrostatic equilibrium.  Transonic motion may be sufficient
to account for what is seen.

Unfortunately we cannot distinguish between a thermal and non-thermal
source of the diffuse emission on spectral grounds, nor identify an
iron line.

\subsubsection{Shocks and mergers}
Alternatively, if the central engine underwent a massive explosion in
the past we may be observing the shocked material behind two shock
fronts travelling to the north and south. In that case the AGN must be
embedded in a dense environment such as a cluster, as is shown by the
calculated electron density. A factor of 4 density increase by the
shock would imply $n_e \sim 7 \times 10^{-3} \pcmcu$ in the preshocked
gas, unless adiabatic expansion was already strong. The preshock
temperature is unknown but must be such that the gas outside the shock
is undetectable (e.g. $\mathrm{k}T<1.5$~keV). The total thermal energy
in the gas is $\sim 1.5 \times 10^{61} \erg$. The shock velocity must
be greater than $2000 \kmps$.  Therefore the time since the explosion
is $\sim 5 \times 10^7 \yr$.  The total energy in the expanding gas
(thermal plus kinetic) must be $\sim 2 \times 10^{61} \erg$, and the
power greater than $10^{46} \ergps$.

Simpson \& Rawlings (2002) make the hypothesis that the diffuse X-ray
emission in 3C294 is produced by two shocks propagating in opposite
directions from the site of a collision of two clusters. In this model
the powerful radio source in the core of 3C294 would have been
triggered by the merger. If the collision velocity were $\sim
1000-2000 \kmps$ then there would be enough energy to raise the
temperature of the cluster from 2 to 5~keV. One problem with this
model is that the surface brightness profile we observe does not match
that expected from simulations of merging galaxies (e.g. Ritchie \&
Thomas 2002). The region of emission is not expected to have a flat
brightness profile and the edges of the heated region are not sharp.

It is interesting that although the diffuse emission from 3C294 does
not appear to have the same morphology as some clusters of galaxies
thought to be merging (e.g. Markevitch \& Vikhlinin 2001), there is a
resemblance with Abell~2256 which has subcomponents of $\sim 4.5$~keV
and $7-8$~keV in an early state of merger at redshift 0.058 (Sun et
al. 2002). The core of A2256 has two sharp and smooth edges like
3C294. However the physical scale of the core is about twice as large
than that of the emission from 3C294, and the profile not as flat.

Carilli et al. (2002) argue that diffuse X-ray emission seen around
the narrow-line radio galaxy PKS~1138-262 is caused by the radio
source shocking a cocoon of gas. The morphology of the X-ray emission
from 3C294 suggests this is not the case in this object, as the
emission does not have the ellipsoidal shape of a radio cocoon (e.g.
Cygnus-A, Smith et al. 2002).

In summary, the evidence suggests that a thermal origin for the
diffuse emission is implausible unless the gas is in a dynamically
active state, and requires significant gas extending beyond that seen.

\subsection{Non-thermal}
A different interpretation is that we are observing X-rays from the
central AGN scattered by the electrons in a surrounding cluster (see
Sazonov, Sunyaev \& Cramphorn 2002). In this model the central source
switched on only recently, and the flux rise time was short so the
furthest edges are sharp.  A toroidal dust cloud around the nucleus
could collimate the radiation into the two cones seen to either side
of the nucleus. This model has difficulty in producing the observed
surface brightness profile. Since the flux drops as the square of the
radius, and the density of any plausible scattering medium will
decrease with distance from the AGN, the brightest part of the
scattered radiation should be closest to the AGN, with the brightness
declining quickly with radius.

\begin{figure}
  \centering
  \includegraphics[angle=270,width=0.99\columnwidth]{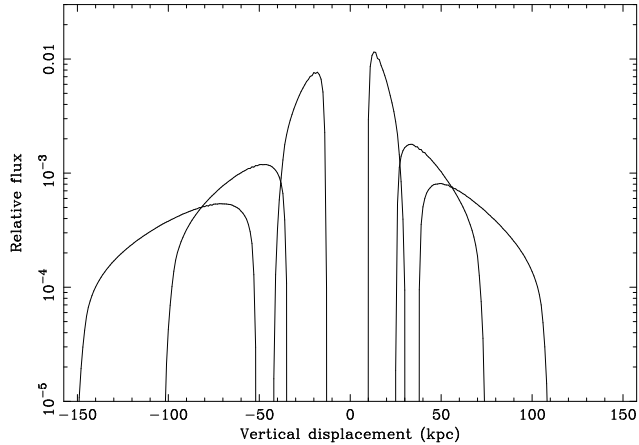}
  \caption{Surface brightness profiles from the simple scattering
    model. Three times are shown, $7 \times 10^4$, $1.7 \times 10^5$,
    and $2.5 \times 10^5$~yr, from the innermost out.}
  \label{fig:scatter}
\end{figure}

The brightness profile can be flattened by inclining the radiation
cone towards the observer. To test whether this is a viable
explanation for the observed surface brightness profile, we made a
simple numerical scattering model. We assumed the optical depth was
low enough so that only single scattering events were important. The
emission along a line of sight was calculated by integrating along
that line the product of the density of the scattering material (taken
to be proportional to the reciprocal of the distance from the centre),
and the flux at that position at the time when the observer sees it
(originally emitted from a central source, but taking into account
light travel time). The emission is integrated along positions which
lie inside the radiation cones. The cones have an opening angle of
$30^{\circ}$ and are inclined at an angle of $15^{\circ}$ towards and
away from the observer, rotated about the $x$-axis. The central source
was switched on at $t=0$ for $10^4$~yr.  The emission along each line
of sight was then integrated along the $x$-axis to simulate the
surface brightness cut in Fig.~\ref{fig:surbrightnorth}.

We show in Fig.~\ref{fig:scatter} the calculated profiles for three
different times. Although the profiles appear fairly flat, in linear
terms they fall steeply from the centre. There is also a gap in the
centre, which is caused by the growing shell of photons moving along
the cones. The gap may be avoided by modifying the burst of radiation
from the core to have a longer-living tail. 3C294 has a dip in surface
brightness to the north of the core, but it is shallow. This model has
difficulty in reproducing the flat brightness profile across the
cluster. If the density profile in the cluster were inverted (as for a
bubble, see Section \ref{sect:hybrid}), or if the radiation was
confined close to the plane of the sky in a fan-like shape, then the
brightness profile could be flattened.

A further problem with this model is that only a small fraction of the
photons emitted will be scattered by the cluster. We can estimate what
the required power for the central source is. The luminosity of the
cones (excluding the radio axis) is $\sim 3 \times 10^{44} \ergps$. If
a third of the sky is subtended by the cones (approximately correct if
the opening angles of the cones are $30^{\circ}$), and 1 per cent of
the radiation is scattered by the electrons, then this would require
an X-ray luminosity between 1 and 10~keV from the central source of
$\sim 10^{47} \ergps$.  The total bolometric power of the source would
be significantly larger.  This luminosity is much higher than that
seen (Section \ref{sect:core}) but the source may be variable. It
would need to be powered by accretion onto a $10^{10}\Msun$ black
hole.

A different non-thermal model is that we are observing inverse Compton
scattered radiation from relativistic electrons. In that case the
electrons could either be a population scattering CMB (Cosmic
Microwave Background) photons requiring energies of $\gamma \sim 1000$
(in order to scatter photons into the 1~keV band), or $\gamma \sim
10-100$ to scatter UV-IR radiation from the central source (e.g.
Brunetti, Setti \& Comastri 1997).  The energy density in the CMB at
that redshift was $\sim 2 \times 10^{-11} \ergpcmsq$, and the energy
density from the nucleus would be $\sim 3 \times 10^{-12} \: L_{47} \:
d_{2}^{-2} \ergpcmsq$ ($L_{47}$ is the luminosity of the source in
units of $10^{47} \ergps$, and $d_2$ is the distance from the source
in units of $10^2 \kpc$). The inverse Compton cooling time of the
CMB-scattering electrons would be $\sim 6 \times 10^7 \yr$ and the
cooling time of the UV-scattering electrons longer by a factor of
$\sim 1000$. We therefore require the total energy in relativistic
electrons to be $\sim 5 \times 10^{59} \erg$ if the radiation was
scattered from the CMB, and a factor 1000 times greater if it was
scattered from the UV radiation field. We now ignore the UV case given
its energy requirements. The total energy in relativistic particles
may be 10 to 100 times larger than $5 \times 10^{59} \erg$ if we
account for other particles such as low energy electrons and protons.
To provide an equivalent energy density to the CMB magnetically would
require a field of $24 \: \mu$G, which is much more than that found in
low-redshift clusters ($B \sim \textrm{few}\: \mu$G; Clarke, Kronberg
\& B\"ohringer 2001).  Providing the magnetic field is of a similar
strength to that in low redshift clusters, inverse Compton losses are
expected to dominate over synchrotron losses.  $\gamma \sim 1000$
electrons will radiate at around $2 \MHz$ in a $\mu$G field, but
otherwise would be undetectable. Such an `inverse Compton cloud' model
would be the most energetically efficient way of explaining the data.
In order to confine this relativistic plasma the gaseous medium of at
least a group would be required in order not to lose the energy by
adiabatic expansion.

An inverse Compton origin for the diffuse X-rays surrounding 3C294
therefore appears to be energetically feasible.  Such a component in
lower redshift clusters will not be dominant, owing to the sixty-fold
lower energy density of the CMB. It may however be detectable (e.g.
the Coma cluster, Fusco-Femiano et al. 1999; see also the theoretical
discussion by Sarazin 1999). Inverse Compton scattering is also
invoked to explain the hotspots in several other radio sources
(Celotti, Ghisellini \& Chiaberge 2001; Hardcastle, Birkinshaw \&
Worrall 2001), and may account for the excess emission from the radio
hotspots in 3C294 (Section \ref{sect:radio_hotspots}). The X-ray
morphology of the diffuse emission does not match that of the radio
source (Fig.  \ref{fig:radio}; McCarthy et al. 1990) or of deeper
radio images (K. Blundell, priv. comm.). In particular, the X-ray
emission to the NW and SE of the nucleus has no radio counterpart.
That need not be a major problem though, since the higher energy
electron population ($\gamma \gtrsim 10^4$) necessary for radio
emission would have a lifetime of $<10^6$~yr and so be absent. We
could simply be seeing the older electron population in a source where
the jet direction has changed by $\sim 60^{\circ}$ over the past
$10^8$~yr.

\subsection{A hybrid model}
\label{sect:hybrid}
Many clusters seen at low redshift with central radio sources have
bubbles of radio plasma displacing the X-ray emitting gas (e.g.
Perseus cluster, Fabian et al. 2000; Hydra A, McNamara et al. 2000;
Cygnus A, Wilson, Young \& Shopbell 2000). Perhaps the powerful radio
source 3C294 has created exceptionally large bubbles in the NS
direction in the surrounding ICM on a timescale of $\sim 10^8$~yr. As
for the bubbles in Hydra A and Perseus, the expansion may be subsonic.
The gas will fall back when the radio source drops in intensity.  The
X-ray emission can then be partially thermal, at large distances from
the centre, and non-thermal nearer the centre, particularly along the
radio axis.

If we are observing bubbles, the pressure of non-thermal particles in
the bubbles will be close to the pressure of the surrounding thermal
gas. If we assume a cooling time for the electrons of $\sim 6 \times
10^7$~yr, emitting a power of $3 \times 10^{44} \ergps$, then the
total energy in the bubbles would be $5.7 \: a \times 10^{57} \erg$
(where $a$ is a factor to account for energy in particles not emitted
in the band we are observing in). Assuming a volume of $4 \times
10^{70} \cmcu$, the pressure inside the lobes would be $1.5 \: a
\times 10^{-11} \ergpcmcu$. This pressure is consistent with the
pressure found in many clusters.

In this picture, were 3C294 to be at $z=0$, the inverse Compton
emission would be 60 times weaker, and thermal emission from the
surrounding cluster would dominate the X-ray image. The bubbles could
appear as holes in the X-ray emission.

\section{Conclusions and Discussion}
3C294 is a powerful, type II, radio-loud quasar, surrounded by
extensive diffuse X-ray emission. A diffuse ICM is likely to be
present, at least as the working surface for the radio lobes and
providing at least some the X-ray emission via bremsstrahlung. The
sharp N and S edges to the X-ray emission mean, for a thermal origin,
that the gas is not in complete hydrostatic equilibrium. It may have
been displaced within the inner third by a bubble of relativistic
plasma. Inverse Compton emission from relativistic electrons with
$\gamma \sim 10^3$ probably accounts for emission from the radio
hotspots, and possibly all the diffuse emission if such a population
exists well beyond the radio structure. The X-ray spectrum of the
diffuse emission mildly favours a power-law of non-thermal origin,
provided that the preliminary \textsc{acisabs} model for the
correction of the soft X-ray degradation of the detector is adequate.

3C294 is an exceptional object, and may be revealing an energetic
phase relevant to many rich clusters of galaxies in which a powerful
active nucleus transfers considerable energy, and a relativistic
particle population, into the ICM. This may be crucial for explaining
the properties of present-day clusters (see e.g. Wu, Fabian \& Nulsen
2000).

\section*{Acknowledgements}
ACF and CSC thank the Royal Society for their support. The authors are
also grateful to Poshak Gandhi for his analysis of the available
optical data for the NE source.

The \emph{HST} data presented in this paper were obtained from the
Multimission Archive at the Space Telescope Science Institute (MAST).
STScI is operated by the Association of Universities for Research in
Astronomy, Inc., under NASA contract NAS5-26555.

\end{document}

%% file: defn.tex
\newcommand{\Mpc}{\rm\thinspace Mpc}
\newcommand{\kpc}{\rm\thinspace kpc}

\newcommand{\km}{\rm\thinspace km}

\newcommand{\cm}{\rm\thinspace cm}

\newcommand{\cmcu}{\hbox{$\cm^3\,$}}
\newcommand{\pcmcu}{\hbox{$\cm^{-3}\,$}}

\newcommand{\yr}{\rm\thinspace yr}

\newcommand{\s}{\rm\thinspace s}

\newcommand{\MHz}{\rm\thinspace MHz}

\newcommand{\Msun}{\hbox{$\rm\thinspace M_{\odot}$}}

\newcommand{\erg}{\rm\thinspace erg}

\newcommand{\ergpcmsq}{\hbox{$\erg\cm^{-2}\,$}}
\newcommand{\ergpcmcu}{\hbox{$\erg\cm^{-3}\,$}}
\newcommand{\ergpcmsqps}{\hbox{$\erg\cm^{-2}\s^{-1}\,$}}

\newcommand{\ergps}{\hbox{$\erg\s^{-1}\,$}}

\newcommand{\kmps}{\hbox{$\km\s^{-1}\,$}}

\newcommand{\kmpspMpc}{\hbox{$\kmps\Mpc^{-1}$}}

\newcommand{\Zsun}{\hbox{$\thinspace \mathrm{Z}_{\odot}$}}

\newcommand{\psqcm}{\hbox{$\cm^{-2}\,$}}
\newcommand{\pcmsq}{\hbox{$\cm^{-2}\,$}}

\newcommand{\ps}{\hbox{$\s^{-1}\,$}}

\newcommand{\amin}{\rm\thinspace arcmin}
\newcommand{\aminsq}{\hbox{$\amin^{2}\,$}}
\newcommand{\asec}{\rm\thinspace arcsec}
\newcommand{\asecsq}{\hbox{$\asec^{2}\,$}}
\newcommand{\kpcsq}{\hbox{$\kpc^{2}\,$}}

